\documentstyle[prl,aps]{revtex}
\input epsf
\begin{document}

\twocolumn[\hsize\textwidth\columnwidth\hsize\csname
@twocolumnfalse\endcsname

\draft

\title{Carbon clusters near the crossover to fullerene stability}

\author{P.~R.~C.~Kent, M.~D.~Towler, R.~J.~Needs, and G.~Rajagopal}

\address{Cavendish Laboratory, Madingley Road, Cambridge CB3 0HE, UK}

\date{\today}
\maketitle

\begin{abstract}
\begin{quote}
\parbox{16 cm}{\small

 The thermodynamic stability of structural isomers of
$\mathrm{C}_{24}$, $\mathrm{C}_{26}$, $\mathrm{C}_{28}$ and
$\mathrm{C}_{32}$, including fullerenes, is studied using density
functional and quantum Monte Carlo methods.  The energetic ordering of
the different isomers depends sensitively on the treatment of electron
correlation.  Fixed-node diffusion quantum Monte Carlo calculations
predict that a $\mathrm{C}_{24}$ isomer is the smallest stable
graphitic fragment and that the smallest stable fullerenes are the
$\mathrm{C}_{26}$ and $\mathrm{C}_{28}$ clusters with
$\mathrm{C}_{2v}$ and $\mathrm{T}_{d}$ symmetry, respectively. These
results support proposals that a $\mathrm{C}_{28}$ solid could be
synthesized by cluster deposition.}
\end{quote}
\end{abstract}

]

\narrowtext 

Since the discovery of the fullerene
$\mathrm{C}_{60}$\onlinecite{HWKrotoNat1985}, the study of carbon
clusters has revealed a rich variety of physical and chemical
properties.  Fullerene clusters may now be synthesised in macroscopic
quantities, but despite many experimental and theoretical advances the
detailed energetics of these systems are not yet fully understood. The
question ``which is the smallest stable fullerene?'' remains both
interesting and contentious due to the sensitivity of cluster
formation to experimental conditions and the challenges posed to
theoretical methods by system size and the high accuracy required. In
this Letter we report very accurate calculations of the relative
energies of $\mathrm{C}_{24}$, $\mathrm{C}_{26}$, $\mathrm{C}_{28}$
and $\mathrm{C}_{32}$ clusters, and identify the smallest stable
fullerenes.

The number of low-energy candidate structures can be large, even
for quite small clusters, precluding exhaustive theoretical searches
with highly accurate but computationally expensive methods. In
practice, a hierarchy of methods of increasing accuracy and
computational cost must be used. The first step is
to select candidate structural isomers via empirical methods based on
bond counting and geometric ``rules'' such as ``minimize the
number of adjacent pentagons''\onlinecite{HWKrotoNat1987}. Quantum
mechanical calculations based on tight-binding and density functional
theory (DFT) methods can then used to refine the selection. To finally
establish the energetic ordering of different isomers, highly accurate
calculations must be performed.  Quantum chemical methods, such as
coupled cluster (CC) calculations\onlinecite{JCizekACP1969}, are
potentially highly accurate, but are severely limited by the size of
basis set that is computationally affordable in these systems.
Quantum Monte Carlo (QMC) methods give an accurate treatment of
electron correlation which, combined with an absence of basis set
error, favorable scaling with system size and suitability for
parallel computation, renders them ideal for these studies.  QMC
calculations have reproduced experimental binding energies of small
hydrocarbons to within 1\%\onlinecite{JCGrossmanPRL1995}. Using the
techniques described below we have calculated the cohesive energy of
bulk diamond, obtaining values of 7.36(1) and 7.46(1) eV
per atom in variational Monte Carlo (VMC) and diffusion Monte Carlo
(DMC), respectively, which are in very good agreement with the
experimental value of 7.37 eV.

Carbon clusters are particularly challenging to model accurately due
to the wide range of geometries and the occurrence of single, double,
and triple bonds.  These differences result in a non-cancelation of
errors in relative energies, exaggerating any errors due to
approximations involved in electronic structure methods. Despite these
potential difficulties, carbon clusters have been extensively studied
using methods such as tight-binding, density functional, quantum
chemical and
QMC\onlinecite{JCGrossmanPRL1995,fullerenecalcbook,GEScuseriaSci1996,AVOrdenCR1998,YShlakhterJCP1999}.
The need for high accuracy calculations with a sophisticated treatment
of electron correlation has been clearly illustrated by several
previous studies.  Grossman {\em et
al.}~\onlinecite{JCGrossmanPRL1995} have performed diffusion Monte
Carlo calculations for $\mathrm{C}_{20}$ clusters, finding that the
fullerene is not energetically stable. A DFT study of
$\mathrm{C}_{20}$ isomers showed that the local density approximation
(LDA)~\onlinecite{JPPerdewPRB1981} and the BLYP gradient-corrected
functional~\onlinecite{BLYP} gave different energy orderings for the
sheet, bowl and fullerene structures, with neither agreeing with that
of DMC\onlinecite{JCGrossmanPRL1995}. Jensen {\em et
al.}\onlinecite{FJensenJCP1998} made calculations for the monocyclic
ring and fullerene isomers of $\mathrm{C}_{24}$ which demonstrated
significant differences between the predictions of the LDA, gradient
corrected and hybrid density functionals.  These authors also
performed second-order M$\o$ller-Plesset and CC calculations, but with
a limited basis set,
concluding that the fullerene is lower in energy than the ring.
Raghavachari {\em et al.}~\onlinecite{KRaghavachariCPL1994} and Martin
{\em et al.}~\onlinecite{JMLMartinCPL1996b} studied seven isomers of
$\mathrm{C}_{24}$ using DFT, but obtained conflicting energetic
orderings.

For clusters containing between 20 and 32 atoms, three classes of
isomer are energetically competitive: fullerenes, planar or
near-planar sheets and bowls, and monocyclic rings. The smallest
possible fullerene, defined as a closed cage containing only
pentagonal and hexagonal faces\onlinecite{PWFowlerAtlas}, consists of
20 atoms. However, the smallest fullerenes most commonly identified by
time of flight and mass spectroscopy measurements are the
$\mathrm{C}_{30}$ and $\mathrm{C}_{32}$
clusters\onlinecite{HKietzmannPRL1998,HHandschuhPRL1995,GvonHeldenCPL1993}.
Rings are found to dominate up to approximately 28 carbon atoms under typical
experimental conditions, and fullerenes are mostly observed for larger
clusters, although other structures are also present (see
for example, Ref.~\onlinecite{GvonHeldenCPL1993}). In this work we
present a QMC study of five isomers of $\mathrm{C}_{24}$, three of
$\mathrm{C}_{26}$ and $\mathrm{C}_{28}$, and two of $\mathrm{C}_{32}$,
thereby covering the range of masses from where the fullerene is
clearly predicted to be unstable to where a fullerene is clearly
observed. This enables us to predict the smallest energetically
stable fullerene.

We apply the diffusion quantum Monte Carlo
method\onlinecite{dmc,BLHammond94} in which the imaginary time
Schr\"{o}dinger equation is used to evolve an ensemble of electronic
configurations towards the ground state.  The ``fixed node
approximation'' is central to this method; the nodal surface of the
exact fermionic wave function is approximated by that of a guiding
wave function.  Core electrons were modeled by an accurate
norm-conserving pseudopotential\onlinecite{NTroullierPRB1991},
and the non-local energy was evaluated stochastically within the locality
approximation\onlinecite{nonlocaleval}. We used Slater-Jastrow guiding
wave functions consisting of the product of a sum of Slater
determinants of single-particle orbitals obtained from
CRYSTAL95\onlinecite{CRYSTAL95} or Gaussian94\onlinecite{Gaussian94}
with a Jastrow correlation
factor\onlinecite{AJWilliamsonPRB1996}. Optimized uncontracted valence
Gaussian basis sets of four \emph{s}, four \emph{p} and one \emph{d}
function were used to represent the single-particle orbitals.  Jastrow
factors of up to 80 parameters were optimized using efficient variance
minimization techniques\onlinecite{CJUmrigarPRL1988,PRCKentPRB1999},
yielding 75-90\% of the DMC correlation energy.

\begin{figure}
\epsfxsize=8cm \epsfbox{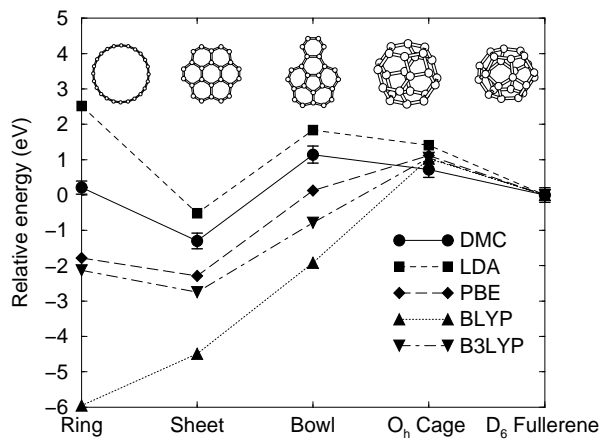}
\caption{The structures and energies of the $\mathrm{C}_{24}$ isomers
given relative to the $\mathrm{D}_6$ fullerene.}
\label{c24graph}
\end{figure}

We relaxed the structures by performing highly converged density
functional calculations.  The geometries were obtained from
all-electron calculations\onlinecite{Gaussian94} using the B3LYP
hybrid functional\onlinecite{ADBeckeJCP1993} and Dunning's cc-pVDZ
basis set~\onlinecite{THDunningJCP1989}, which has been found to be an
accurate and affordable
combination\onlinecite{JMLMartinCPL1996b,JMLMartinMP1995,JMLMartinCPL1996a}.
To assess the sensitivity of the total energies to the geometries, we
compared the energies of the fully relaxed ring and $\mathrm{D}_6$
fullerene isomers of $\mathrm{C}_{24}$ (see Fig.~\ref{c24graph}) using
the BLYP and B3LYP functionals. The functionals give significantly
different energetic orderings, but the differences between the
geometries are small - less than 0.03 angstroms in bond lengths and
0.4 degrees in bond angles.  The relative energies of these structures
changed by a maximum of 0.27 eV for each of the functionals
investigated.  The relative energies are therefore rather insensitive
to the functional used to obtain the geometries, but are more
sensitive to the functional used to calculate the energies.  These
changes are small compared with the overall range of energies, but
some changes in the orderings of the isomers closest in energy could
occur.

We considered the following isomers of $\mathrm{C}_{24}$, as depicted
in Fig.~\ref{c24graph}: a polyacetylenic monocyclic ring, a flat
graphitic sheet, a bowl-shaped structure with one pentagon, a caged
structure with a mixture of square, pentagonal and hexagonal faces,
and a fullerene.  Other candidate structures, such as bicyclic rings
and a 3-pentagon bowl were excluded on the grounds that DFT
calculations using several different functionals have shown them to be
significantly higher in
energy\onlinecite{KRaghavachariCPL1994,JMLMartinCPL1996b}.  As well as
DMC calculations we have also performed DFT calculations using the
LDA, two gradient corrected functionals
(PBE\onlinecite{JPPerdewPRL1996} and BLYP) and the B3LYP functional.
The results shown in Fig.~\ref{c24graph} confirm that the treatment of
electron correlation has a profound effect on the relative energies.
All of the functionals give different energetic orderings, and none
gives the same ordering as DMC.  The graphitic sheet is placed lowest
in energy by DMC, in agreement with each of the functionals except
BLYP, which places the ring lowest in energy.  The low energy of the
$\mathrm{C}_{24}$ graphitic sheet is expected because the structure
accommodates a large number (7) of hexagonal rings without significant
strain.  This structure is predicted to be the smallest stable
graphitic fragment. Both DMC and the DFT approaches find the
$\mathrm{C}_{24}$ fullerene to be unstable.

\begin{figure}
\epsfxsize=8cm \epsfbox{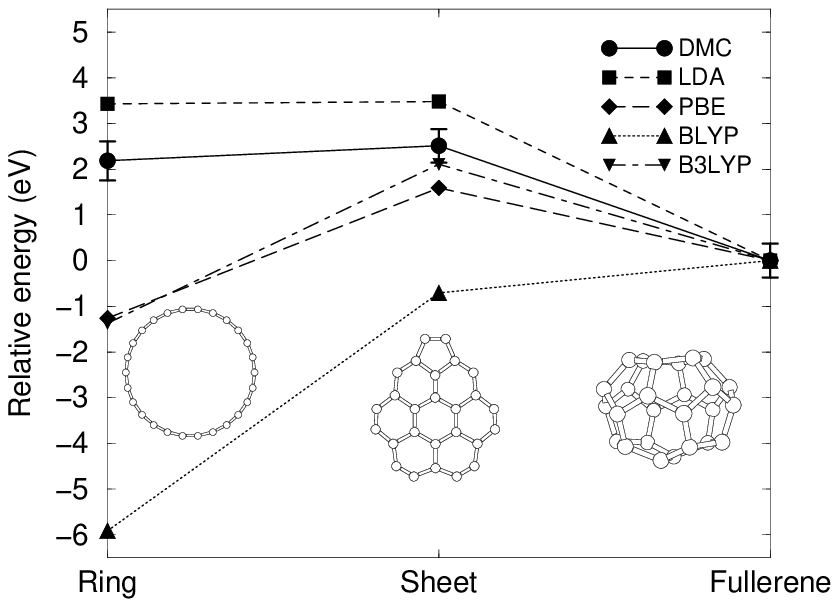}
\caption{The structures and energies of the $\mathrm{C}_{26}$ isomers,
given relative to the fullerene.}
\label{c26graph}
\end{figure}

Three isomers of $\mathrm{C}_{26}$ were considered: a cumulenic
monocyclic ring, a graphitic sheet with one pentagon and a fullerene
of $\mathrm{C}_{2v}$ symmetry (Fig.~\ref{c26graph}). Few studies of
the $\mathrm{C}_{26}$ fullerene have been made, in part due to the
high strains evident in its
structure\onlinecite{PWFowlerAtlas}. Recently Torelli and
Mit\'{a}\v{s} have demonstrated the importance of using
multi-determinant trial wave functions to describe aromaticity in 4N+2
carbon rings~\onlinecite{TorelliUnpub}.  We have tested this for the
$\mathrm{C}_{26}$ ring, using a 43 determinant trial wave function
obtained from a CI singles-doubles calculation.  The multi-determinant
wave function gave a slightly lower DMC energy than the single
determinant wave function, by approximately $0.5$ eV, confirming that
CI wave functions can have better nodal surfaces than HF wave
functions. The ring and sheet-like isomers are close in energy, but
the fullerene is approximately $2.5$ eV below these isomers and is
therefore predicted to be the smallest stable fullerene. Small changes
in the geometries are highly unlikely to change this conclusion.

\begin{figure}
\epsfxsize=8cm \epsfbox{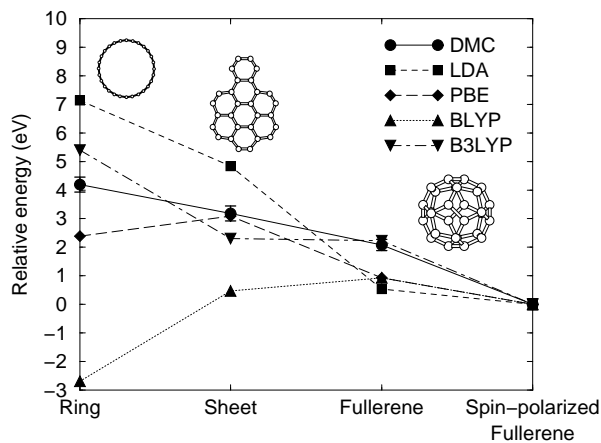}
\caption{The structures and energies of the $\mathrm{C}_{28}$ isomers,
given relative to the spin-polarized fullerene (see text).}
\label{c28graph}
\end{figure}

Three $\mathrm{C}_{28}$ isomers were investigated: a monocyclic ring,
a graphitic sheet and a fullerene of $\mathrm{T}_{d}$ symmetry
(Fig.~\ref{c28graph}).  Other bowl and sheet-like structures were
excluded on energetic grounds\onlinecite{JMLMartinCPL1996a}.
Spin-polarized DFT calculations show the ground state of the
$\mathrm{T}_{d}$ symmetry fullerene to be a spin-polarized $^5A_2$
state.  DMC predicts that this spin-polarized fullerene is the lowest
energy isomer of $\mathrm{C}_{28}$, and this is supported by each of
the functionals except BLYP.  The spin-polarized fullerene has four
unpaired electrons and is therefore highly reactive.  This property
has been exploited in atom trapping experiments in which fullerenes
containing single four-valent atoms, $\mathrm{C}_{28}\mathrm{M}$, have
been prepared by laser vaporization of a graphite-MO$_2$ (M = Ti, Zr,
Hf or U) composite rod\onlinecite{TGuoSci1992}. Our prediction that
the fullerene is the most stable isomer of $\mathrm{C}_{28}$ indicates
that isolated fullerenes might be readily produced. This would
facilitate investigations of $\mathrm{C}_{28}$ fullerene solids, which
have been discussed but not yet
produced\onlinecite{TGuoSci1992,JKimJCP1998}, although this route may
be hampered by the chemical reactivity of the 
fullerene. (A $\mathrm{C}_{36}$ fullerene solid has been
reported\onlinecite{CPiscottiNat1998}.)

Our DFT and DMC results for $\mathrm{C}_{28}$ (Fig.~\ref{c28graph})
again highlight a wide variation between different DFT functionals.
The LDA and B3LYP functionals predict the same ordering as DMC, but
the PBE and BLYP functionals give different orderings.  The DMC data
strongly indicates that the $\mathrm{T}_{d}$ fullerene is the most
stable $\mathrm{C}_{28}$ isomer at zero temperature. The fullerene has
the lowest DMC energy in both spin-polarized and non spin-polarized
calculations, and is substantially more stable than the sheet and
ring.  Small changes in the geometries are therefore unlikely to
change this ordering.

\begin{figure}
\epsfxsize=8cm \epsfbox{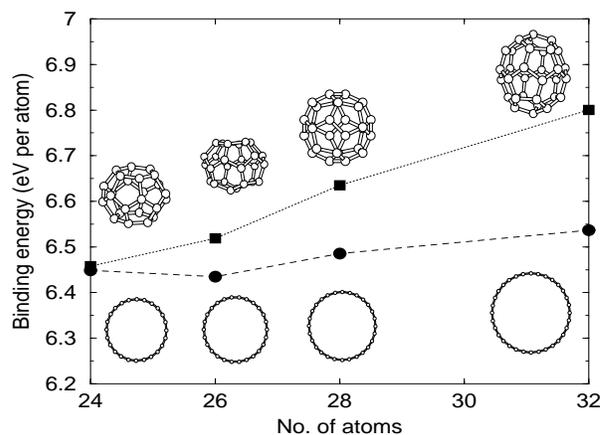}
\caption{The DMC binding energies of the
$\mathrm{C}_{24}$, $\mathrm{C}_{26}$, $\mathrm{C}_{28}$,
$\mathrm{C}_{32}$ ring and fullerene structures. The lines drawn are
for guidance only. Statistical error bars are smaller than the
symbols.}
\label{cbindingfig}
\end{figure}

Our DMC calculations for the $\mathrm{C}_{32}$ monocyclic ring and
fullerene show that the fullerene is 8.4(4) eV per molecule lower in
energy, which is consistent with the observation of a large abundance
of $\mathrm{C}_{32}$ fullerenes in a recent cluster
experiment\onlinecite{HKietzmannPRL1998}. In Fig.~\ref{cbindingfig} we
plot the DMC binding energies per atom of all the ring and
fullerene structures considered.  The binding energies of the fullerenes
rise much more rapidly with cluster size than those of the rings because
of the large amount of strain in the smaller fullerenes.  The
DMC binding energy of the $\mathrm{C}_{32}$ fullerene is
approximately $1$ eV per atom less than the experimental binding energy
of $\mathrm{C}_{60}$.

Our DFT and DMC results highlight several important trends in the relative
performance of the different functionals.  The overall quality of a
functional for the clusters is best judged by the agreement with the
DMC data for the overall {\it shapes} of the relative energy data of
Figs.~\ref{c24graph}-\ref{c28graph}.  The best agreement is given by
the PBE and B3LYP functionals, with the LDA being slightly inferior
and the BLYP functional being worst.  The tendency of the LDA to favor
structures of high average coordination and for the BLYP functional to
favor structures of low average coordination is consistent with
the results for $\mathrm{C}_{20}$ reported by Grossman {\it et
al.}\onlinecite{JCGrossmanPRL1995}.

The final test of our predictions must lie with experiment.  It is
clear that the actual abundances of different clusters depend
sensitively on experimental conditions.  Analysis of the stability of
clusters against fragmentation, growth and other chemical reactions is
complicated.  One key issue is that the clusters are formed at
temperatures of order 1000K and therefore the vibrational
contributions to the free energy can be significant.  Fortunately, a
simple picture emerges from computations of vibrational
properties~\onlinecite{KRaghavachariCPL1994,JMLMartinCPL1996b,JMLMartinCPL1996a}.
Fullerenes are relatively rigid and have higher vibrational free
energies than rings, which have many low-lying vibrational modes.
Vibrational effects therefore tend to favor the ring isomers at high
temperatures.  However, according to our DMC calculations the
$\mathrm{C}_{26}$ and $\mathrm{C}_{28}$ fullerenes are several eV per
cluster lower in energy than the other isomers, so that significant
amounts of fullerene could exist at the temperatures of formation.  If
thermodynamic stability alone were to determine which cluster sizes
were observed then only the largest fullerenes would ever be observed,
but in a recent experiment the abundance of the $\mathrm{C}_{32}$
fullerene was found to be {\em greater} than
$\mathrm{C}_{60}$\onlinecite{HKietzmannPRL1998}. There is more
evidence that thermodynamic stability to rearrangements of clusters of
a particular size are important in determining which isomers are
observed.  For example, in the experimental study of
Ref.~\onlinecite{HKietzmannPRL1998}, fullerenes were mostly observed
for clusters containing more than about 30 carbon atoms, while for
smaller clusters mostly rings were formed.  This crossover is close to
the critical size for fullerene stability of 26-28 atoms predicted by
our DMC calculations.

In conclusion, performing accurate calculations of the relative energies of
carbon clusters is a severe test of electronic structure methods
because of the widely differing geometries and the occurrence of single,
double and triple bonds. In our DMC calculations for
$\mathrm{C}_{24}$, the lowest energy isomer is a graphitic sheet,
which is expected to be the smallest stable graphitic fragment.  We
predict that the smallest energetically stable fullerenes are the
$\mathrm{C}_{2v}$ symmetry $\mathrm{C}_{26}$ cluster and the spin
polarized $^5A_2$ state of the $\mathrm{T}_{d}$ symmetry
$\mathrm{C}_{28}$ cluster. This prediction lends weight to recent
proposals that a $\mathrm{C}_{28}$
solid~\onlinecite{TGuoSci1992,JKimJCP1998} could be synthesized by
surface deposition of $\mathrm{C}_{28}$ fullerenes. 

Financial support was provided by EPSRC (UK). Calculations were
performed on the CRAY-T3E at the University of Manchester and the
Hitachi SR2201 located at the University of Cambridge HPCF.

\end{document}